\documentclass[manuscript]{acmart}
 
\usepackage{xcolor}
\usepackage{fontawesome}
\usepackage{soul}

\usepackage{xspace}
\AtBeginDocument{%
  \providecommand\BibTeX{{%
    \normalfont B\kern-0.5em{\scshape i\kern-0.25em b}\kern-0.8em\TeX}}}

\copyrightyear{2021}
\acmYear{2021}
\setcopyright{acmcopyright}\acmConference[DIS '21]{Designing Interactive Systems Conference 2021}{June 28-July 2, 2021}{Virtual Event, USA}
\acmBooktitle{Designing Interactive Systems Conference 2021 (DIS '21), June 28-July 2, 2021, Virtual Event, USA}
\acmPrice{15.00}
\acmDOI{10.1145/3461778.3462012}
\acmISBN{978-1-4503-8476-6/21/06}

\begin{document}


% User-Data as a Probe for Making AI Design Materials
%Towards a Human-Centered Approach to Co-Creating AI Design Materials
%Towards a Human-Centered Approach to Prototyping/Specifying/Realizing AI Design Materials
\title{Towards A Process Model for Co-Creating AI Experiences}
%\title{Towards a Human-Centered Approach to Co-Creating AI Design Materials}
\author{Hariharan Subramonyam}
\affiliation{%
  \institution{University of Michigan}
  \city{Ann Arbor}
  \country{USA}}
\email{harihars@umich.edu}

\author{Colleen Seifert}
\affiliation{%
  \institution{University of Michigan}
  \city{Ann Arbor}
  \country{USA}}
\email{seifert@umich.edu}

\author{Eytan Adar}
\affiliation{%
  \institution{University of Michigan}
  \city{Ann Arbor}
  \country{USA}}
\email{eadar@umich.edu}

\renewcommand{\shortauthors}{Subramonyam et al.}

\begin{abstract} 
Thinking of technology as a design \textit{material} is appealing. It encourages designers to explore the material's properties to understand its capabilities and limitations---a prerequisite to generative design thinking. However, as a material, AI resists this approach because its properties only \textit{emerge} as part of the user experience design. Therefore, designers and AI engineers must collaborate in new ways to create both the material and its application experience. We investigate the \textit{co-creation} process through a design study with 10 pairs of designers and engineers. We find that design `probes' with user data are a useful tool in defining AI materials. Through data probes, designers construct designerly representations of the envisioned AI experience (AIX) to identify desirable AI characteristics. Data probes facilitate divergent design thinking, material testing, and design validation. Based on our findings, we propose a process model for co-creating AIX and offer design considerations for incorporating data probes in AIX design tools.

\end{abstract}

\begin{CCSXML}
<ccs2012>
   <concept>
       <concept_id>10003120.10003123.10010860.10010859</concept_id>
       <concept_desc>Human-centered computing~User centered design</concept_desc>
       <concept_significance>500</concept_significance>
       </concept>
   <concept>
       <concept_id>10010147.10010178</concept_id>
       <concept_desc>Computing methodologies~Artificial intelligence</concept_desc>
       <concept_significance>500</concept_significance>
       </concept>
   <concept>
       <concept_id>10011007.10011074.10011134</concept_id>
       <concept_desc>Software and its engineering~Collaboration in software development</concept_desc>
       <concept_significance>500</concept_significance>
       </concept>
 </ccs2012>
\end{CCSXML}

\ccsdesc[500]{Human-centered computing~User centered design}
\ccsdesc[500]{Computing methodologies~Artificial intelligence}
\ccsdesc[500]{Software and its engineering~Collaboration in software development}

\keywords{Design Materials, Human-Centered AI, User Data}

\begin{teaserfigure}
  \includegraphics[width=\textwidth]{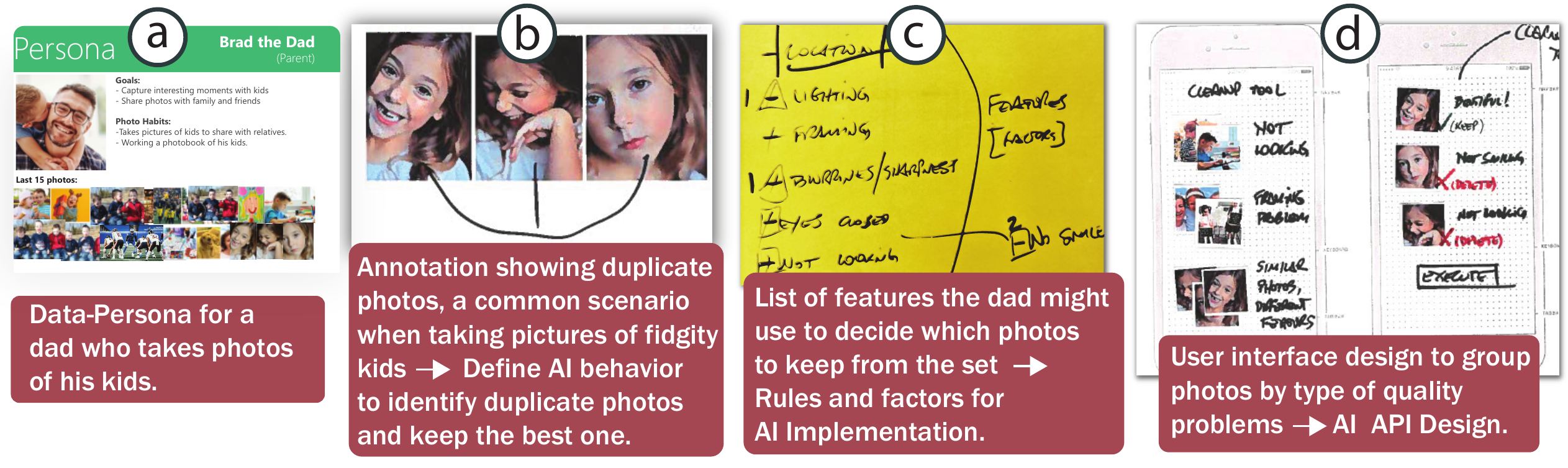}
  \caption{User Data as a probe to design AI material: (a) The designer uses data-persona to construct (b) a scenario about a parent taking duplicate photos (designs AI behavior), (c) listing the features the parent might use to identify bad photos (defines how AI should implement the behavior), and (d) creating AI-Powered UI for de-cluttering photo albums (defines inputs and outputs to AI).}
  \Description{Annotations over printed photos to show created use scenario, list of factors created by participants about mental model, and paper prototype of user interface from session 3 in the study.}
  \label{fig:teaser}
\end{teaserfigure}

\maketitle
\section{Introduction}

When working with new and unfamiliar technology, designers are encouraged to consider it from a ``material'' perspective~\cite{robles2010texturing, wiberg2016interaction, giaccardi2015foundations}. Just as with wood or fabric, in which the craftsman needs to understand the material to create with it, designers need to know what the technology is capable of, what its limitations are, and what properties are available for design. For instance, when working with Radio Frequency ID (RFID) technology, the designer should first explore its material properties including the signal strength, how much information an RF tag can hold, and how quickly the information can be read~\cite{ arnall2014exploring}. This will allow the designer to \textit{manipulate} its properties in creative ways to generate design solutions~\cite{simon1969sciences, council2005double, beaudouin2009prototyping}. However, unlike other technology materials that are created \textit{before} the user experience (UX) design, artificial intelligence (AI) does not lend itself well to a purely material-driven design approach~\cite{karana2015material,zhou2020ml}. Instead, AI's material properties only emerge through its application experience design~\cite{dove2017ux,kayacik2019identifying}. As a simple example, to design an intelligent To-Do List application that automatically creates tasks from emails (e.g.,~\cite{peopleai}), designers cannot work with AI as a given material that makes predictions from text. To create the AI material, AI engineers need guidance from designers about how end-users think about tasks, who the potential end-users are, what emails mean to users, and so on (i.e., human-centered AI~\cite{auernhammer2020human}). For both the designer and the AI engineer, this is a challenging \textit{chicken-and-egg} problem~\cite{seffah2005human, patel2010lowering}. To address this, we investigate what a co-creation process for AI's \textit{form} and \textit{function} might look like, what AI as a material ``under construction'' entails, and how the evolving UX design can inform AI development. 

A fundamental assumption of the material view in HCI is that materials are a given, and they \textit{possess} specific properties that are amenable to design. To a large extent, this assumption holds. From an engineering standpoint, materials are invented with specific \textit{structure-property} relationships in mind (e.g., ~\cite{sutherland1964sketchpad, smith1975pygmalion, jin2020photo}). They can be used in any context in which those relationships are desirable~\cite{alexander1964notes, olson2000designing, heidt2014deconstructivist}. The designer's job is then to explore the material, understand how end-users might experience it, and thereby acquire knowledge for generative design thinking~\cite{giaccardi2015foundations}. For instance, to prototype a To-Do List in mixed-reality (a novel material), the designer can begin by exploring how graphics are rendered spatially, what the visual field of view entails, and which hand gestures are available for interactivity (e.g.,~\cite{nebeling2018protoar}). Based on this, designers can prototype alternatives by changing the appearance of graphical elements~\cite{hololens}, exploring different gestures and layout options to design the UX. In other words, design with material is accomplished by knowing its \textit{created} properties. Even in extreme cases of customization~\cite{leiva2019enact}, the design material metaphor holds. 

In this vein, AI is also a novel design material~\cite{holmquist2017intelligence, yang2018machine}. However, there are important distinctions that make it challenging to put this material perspective into practice. First, as a given (or prefabricated) material, AI is \textit{deficient} for design. AI materials are commonly described in abstractions such as techniques (e.g., supervised-learning) and behavior (e.g., prediction), and divorced from contexts in which the AI is applied. Unlike the mixed-reality example, a designer cannot simply explore a ``supervised learning'' AI to design an intelligent To-Do List. Designing with AI requires \textit{defining} its material properties, including what the AI system should learn using what data, which assumptions and learning rules are appropriate, and how those capabilities should manifest in designed experiences (e.g., data labels). Second, once created by AI engineers, an AI system's properties cannot be readily manipulated during the application's design process. In comparison to mixed reality interfaces, in which the designer can change properties (such as color or shape of elements), representational and knowledge barriers prevent designers from directly altering AI to mold it into a `designed' product (e.g., ~\cite{yang2019sketching}). Third, in many AI systems, its material characteristics can continue to evolve through feedback and learning. The designers must anticipate how the AI will change over time and experience. These capabilities require design across both the application and the AI material.  

To consider the intersection of AI creation and human-centered application design, which we call \textit{AI experience design} (AIX), HCI researchers have put forth design guidelines~\cite{amershi2019guidelines, PAIR}, processes~\cite{zhou2020ml}, and tools~\cite{van2018prototyping}. However, these recommendations emphasize the designer's responsibility to understand the AI design material, but not the role of AI practitioners in AIX. The material design approach assumes the AI, like natural wood, must be taken as given; so, the designer provides the required adaptation. For designers, this challenges their technical expertise and introduces friction to material exploration~\cite{yang2020re}. For instance, designers cannot prototype the user experience with a ``fail fast, fail often~\cite{yang2019sketching}'' approach. With AI material, \textit{vertical} end-to-end prototyping is required to create, evaluate, and revise design alternatives. Such a process is time-consuming for designers and engineers, it is resource-intensive, and the chicken-and-egg problem remains~\cite{kayacik2019identifying}. This is the primary motivation for our work:  \textit{ How might designers and engineers co-create AI Experiences through rapid, collaborative design?}

Drawing from prior research in HCI and AI application design, we developed a protocol for co-creating AIX through generative design thinking. Using the protocol, pairs of UX designers and AI engineers worked to design AI material characteristics and the user experience for a given design problem. We observed UX designers' involvement in AI material creation, including defining AI behavior, specifying the AI architecture, and features related to explainability, failure, and learnability. We found that end-user data played a critical role in shaping AIX. As shown in Figure~\ref{fig:teaser}, by using data as a design probe, designers constructed AI-infused scenarios to co-design desired AI behavior. By imagining mental-models for different personas across scenarios and data (i.e., how might the persona perform a task?), they offered inputs to AI architecture design. Through user interface prototyping with data, participants also co-designed the application programming interface (API). Teams created data probes as a scaffold for divergent design thinking, material testing, and design validation. Our key contributions include (1) identifying the role of designerly proxies\footnote{Designerly proxies are the designer's representations of AI's technical characteristics.} and data probes in defining AIX material, (2) describing a process model for co-creating AIX, and (3) highlighting a set of design considerations for incorporating data probes in AIX design tools.

\section{Related Work}
A material framework for design includes (1) \textit{fabrication}---ways to produce materials with specific properties, (2) \textit{application}---ways to transform materials into products, and (3) \textit{appreciation}---reception of material by the end-users~\cite{doordan2003materials}. Design requires iteration and feedback across these three aspects. When fabrication and application are cleanly separated---as  with natural materials like wood and technological materials like RFID---prior work has considered design material as a \textit{given}~\cite{schon1996reflective,landin2005fragile,hansen2011full,fernaeus2012material,arnall2014exploring,giaccardi2015foundations,dew2018lessons,odom2018design}. As a consequence, UX design emphasizes understanding material \textit{properties}, developing \textit{processes} to generate material artifacts, and evaluating \textit{expectations} and \textit{values} associated with material encounters. Other work has combined material creation (fabrication) into the design problem and investigated \textit{co-creation} of the material along with its application ~\cite{simon1969sciences, redstrom2005technology, blevis2006regarding, vallgaarda2007computational, lindell2012code, jung2012digital}. We draw from both of these perspectives to develop our understanding of designing the AI material and designing with it. By characterizing `design' as an activity that applies a value system to create objects of reasoning~\cite{blevis2006regarding}, with AI as design material, we identify gaps in guidelines, methods, and representations. Through this discussion, we formulate the research questions for our study. 

\subsection{Guidelines} 
Numerous design guidelines for AI applications have emerged from both academic and industry research. The guidelines span across functionality~\cite{Appleguidelines}, end-user interactions~\cite{heer2019agency, amershi2019guidelines, PAIR}, learnability~\cite{gil2019towards}, explainability ~\cite{wang2019designing}, privacy~\cite{jobin2019global, hagendorff2019ethics}, transparency~\cite{eiband2018bringing}, etc. Several guidelines address the intersections of both fabrication and application design (i.e., AIX). In some cases, the guidelines ask that we consider application context when creating AI capabilities; For instance, PAIR~\cite{PAIR} recommends modeling AI after the human expert: \textit{``When designing automation, we should consider how a theoretical human `expert' might perform the task today''}. Others offer suggestions for repairing AI material flaws through UX enhancements: \textit{``Make it easy to edit, refine, or recover when the AI system is wrong~\cite{amershi2019guidelines}.''} This highlights the inherent dependencies between fabrication and application design for AI~\cite{dove2017ux, yang2020re}. In our work, we look at how designers and AI engineers conceptualize the guidelines from different ``points of view'' to co-create AIX. To aid our investigation, we consider the language of material engineering which provides a vocabulary for material as \textit{structure}, \textit{surface}, and \textit{properties}~\cite{vallgaarda2007computational}. This would allow us to establish connections between material characteristics and AI material experience in its embodiment, encounters, and collaborations ~\cite{giaccardi2015foundations}. 

\textit{\textbf{Research Question 1:} How might designers and AI engineers conceptualize shared, and differing, design perspectives arising from human-AI guidelines to co-create AIX?}

\subsection{Design Methods}
Current AI development workflows consist of critical design-related steps, including identifying model requirements, data labeling, feature engineering, etc.~\cite{amershi2019software}. However, in many cases, UX design and AI development only converge after AI decisions have been made~\cite{dove2017ux}. Consequently, UX designers face challenges in incorporating AI material properties within their design practices~\cite{yang2018machine}. Similarly, engineers find it challenging to obtain ground truth validation for AI-related decisions~\cite{hill2016trials}, avoid blind-spots threatening responsible AI needs~\cite{holstein2019improving}, and incorporate necessary UX inputs for improving model performance~\cite{tata2017quick}. Hence, design methodologies should be symbiotic to produce the best application performance. If AI is meant to replicate human intelligence, UX designers can offer insights to make it practically and emotionally resonant with users~\cite{ceconello2019design,yang2017role}. The main challenges to collaboration have been time-related constraints to fabrication and design~\cite{yang2018investigating}, barriers to immediate feedback~\cite{yang2019sketching,kayacik2019identifying}, and lack of motivation and incentives~\cite{kayacik2019identifying,cramer2019confronting}. Further, AI material challenges conventional prototyping methods because it requires a higher level of commitment and effort to prototype AI applications~\cite{yang2020re,yang2019sketching}. We also lack means for UX designers to engage in a ``conversation with the materials~\cite{yang2018machine},'' and ways for the AI materials to ``talk back to the designer~\cite{yang2018machine}.'' 

When considering software code as design material, programming becomes a vital part of the design process and offers necessary ``talk-backs'' for design~\cite{lindell2012code}. Even in natural materials such as wood, material properties (hardness, grain), and constraints (knots and weak points) offer feedback resulting in design recourse~\cite{dew2018lessons}. In co-creating AIX, both designers and engineers require the material and the application experience to respond to each other. Separately, when working with Bluetooth as novel design material, designers generate end-to-end fully working sketches (inspirational bits), allowing them to investigate its properties through form-giving~\cite{sundstrom2011sketching}. To co-create AIX, designers and engineers need similar low-cost \textit{vertical-prototyping} strategies to create end-to-end prototypes of the UX and the AI backend~\cite{beaudouin2009prototyping}. Tools for material prototyping should be accessible (allow developers and designers to think about the material), immediate (support rapid iterative feedback, reflection-in-action, and reflection-on-action), and generative (allow test, probe, and exploration iterations)~\cite{hansen2011full}. Lastly, developers, engineers, and data scientists employ more mechanical, less organic processes focused on application \textit{data}~\cite{girardin2017user}. Consequently, a key task for human-centered designers is promoting user experience data as a bridge between the two fields through a process of translation~\cite{girardin2017user, helms2018design}. We incorporate these perspectives about \textit{data} in developing our study protocol for AIX.

\textit{\textbf{Research Question 2:} How might designers and engineers co-create the design and technical characteristics of AIX?}

\subsection{Representations}
Design requires creating and comparing alternatives to arrive at a final solution~\cite{simon1969sciences}. The challenge with AIX is finding intermediate design representations that can serve as a ``lingua franca'' easily understood by multi-disciplinary teams of designers and engineers~\cite{blevis2006regarding}. Representations would ideally allow the design concept to be viewed differently based on functional perspectives. For instance, for engineers, data is represented as a set of variables~\cite{buitinck2013api}; but in UX, data is associated with end-users and their situated context~\cite{hellerstein2017ground}. Designing across boundaries requires a show and tell (``I will know it when I see it~\cite{boehm2000requirements}'') approach. In~\cite{kayacik2019identifying}, active discussions between UX designers and ML researchers around mock-ups helped them avoid miscommunication about model capabilities. These rich representations should provide ways to envision viewpoints and resolve differences, and  align design needs through negotiation over intermediate representations including words, sketches, physical mock-ups, charts, etc \cite{bucciarelli2002between, huber2019use}. At the same time, these representations should be easy to create during rapid prototyping. In prototyping web applications, designers use non-functional proxies to negotiate a design that works for both developers and end-users (e.g., Interfake~\cite{Interfake}, Apiary~\cite{Apiary}, etc.). Such `mocks' can circumvent the need for more programming effort to creating AI material designs. Our study explores mixed-fidelity prototypes~\cite{mccurdy2006breaking} that provide high-fidelity representations in some dimensions and low fidelity in others.

\textit{\textbf{Research Question 3:} What types of material and design representations can support co-creating AIX?}
\section{Method}
We conducted an in-lab design study in which UX designers paired with AI engineers worked together to co-create AIX (a total of 10 sessions, each with one designer and one engineer). To model the nature of collaboration, we took inspiration from Wizard-of-Oz (WoZ) techniques for AI prototyping~\cite{maulsby1993prototyping, van2018prototyping, browne2019wizard}. We imagined that designers and engineers would implicitly play the `wizard' (experts) role during co-creation. This allows for rapid feedback about both the AI material being created by engineers and application experience being prototyped by the designer. Therefore, we recruited participants who had prior experience in AI application domains and working in collaborative teams; The designers in our study had an average of 3.4 years of experience ($SD=2.8$), and AI engineers had 3.9 years of experience on average ($SD=2.1$). Participants comprised of industry practitioners as well as graduate students with prior work experience (Table ~\ref{table:organization_interviewees}). Participants were paired based on their availability. Each session lasted 2.5 hrs, and we compensated participants with \$40 for their time. All sessions were video-recorded. We collected all artifacts generated by the participants for our analysis. 

\subsection{Study Protocol}
\begin{figure*}[t!]
\centering
\includegraphics[width=\textwidth]{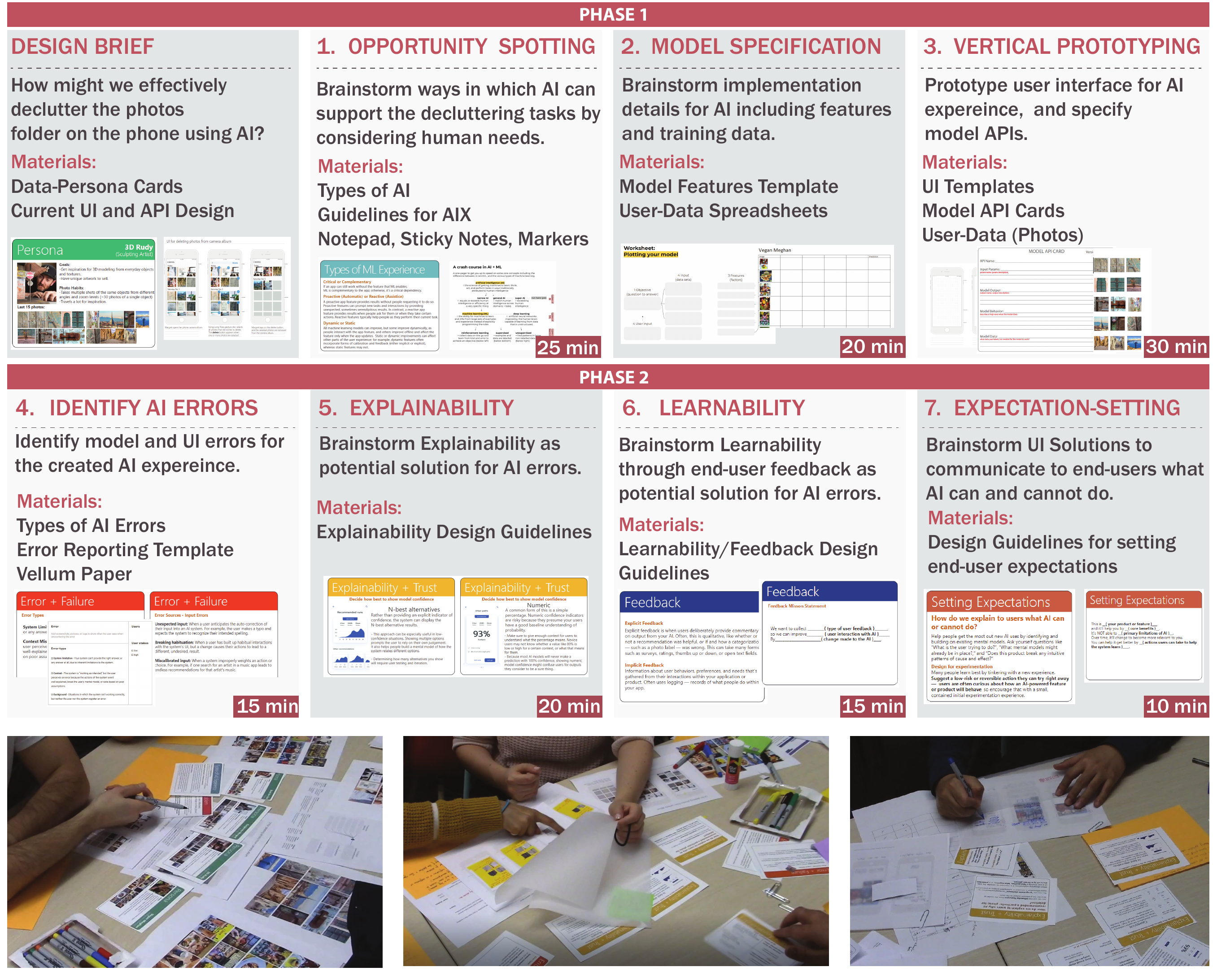}
\caption{Overview of our study protocol for co-creating AIX. Top: Design brief and high-level objective for each of the seven steps. Bottom: Visuals from In-Lab sessions.}
\Description{An infographic showing the design brief and seven steps in the study protocol, and some photos from in-lab sessions showing participants collaborating over different materials. }
\label{fig:protocol}
\end{figure*}

To develop the protocol, we started with human-AI (HAI) design guidelines, developed by companies, for designers and engineers ~\cite{amershi2019guidelines, PAIR, Appleguidelines, aimeetsdesign}. Ideally, these guidelines represent `best-practice' advice that is derived from successful practices within the companies. They offer a starting point to explore how UX and AI roles might collaborate in designing the AI experience. We categorized the guidelines into seven steps spanning AI creation, UX design, and AI-UX design processes (see Figure~\ref{fig:protocol}). Our steps roughly followed the material design process~\cite{karana2015material} of proposing material, envisioning material experience, manifesting material experience patterns, and making material product concept designs. In addition, we included material creation in the process. In our instructions to participants, we refrained from using the material metaphor; instead, we worked with terminology specified in the HAI guidelines. Further, we organized the steps into two phases; The first phase aimed at producing initial AI specifications (fabrication) and prototypes of the AI-powered UI (application). This included opportunity spotting, model specification, and UI prototyping. In the second phase, participants iterated over the design to arrive at a `pragmatic' solution by considering errors, explainability, feedback, and expectation-setting for end-users (appreciation). Our first session served as a pilot informing the two phases, and we used the feedback to revise the protocol for the remaining sessions. The second phase would allow teams to consider AI's uncertainties and offer adaptations to account for AI errors, thereby maximizing AI's utility for end-users. At each step of the protocol, the study coordinator offered instructions and tools for participants to work on that step. Each step had a time limit, and participants shared and discussed their design with the coordinator throughout the study.

\begin{table}[t!]
\small
\centering
\begin{tabular}{@{}l l l@{}}
Session ID & UX Designer & AI Engineer
\\\midrule
1 &  4 months & 1 years \& 4 months\\
2 &  3 months & 6 years \\
3 & 4 years \& 2 months & 2 years \& 1 month\\
4 & 7 years \& 5 months & 2 years \& 9 months\\
5 & 4 years \& 6 months & 1 years \& 6 months\\
6 & 3 months & 7 years \\
7 & 4 years \& 5 months & 2 years\\
8 & 6 years \& 2 months & 4 years \\
9 & 5 years \& 7 months & 5 years \& 6 months\\
10 & 3 months & 6 years \& 5 months\\
\end{tabular}
\caption{Participant details indicating years of experience for designers and engineers in our study.}
\label{table:organization_interviewees}
\end{table}

\textit{Design Problem Briefing:} We selected the problem of \textit{decluttering the photo album} on the phone. We motivated the problem by stating that cameras on smartphones have made it easy to take photos anytime and anywhere. A consequence is that users capture hundreds of photos that may be of little value. Deleting unwanted photos can be tedious and boring. We asked participants to design an AI-powered solution to address this problem. This domain is simple enough for participants to understand, and they can leverage experiential knowledge in brainstorming solutions. In other words, the problem minimizes domain complexities while allowing us to observe collaboration in a single task session. We provided participants with screenshots of the current (non-AI) photo album interface that we created. In our pilot, we observed that the designer spent time creating the same interface by looking at their phone. Providing this design upfront allows us to focus time on more critical steps. We also specified what the back-end delete API looks like (a simple function that takes a list of photo ids to delete and returns a success or failure message). 
 
Like Zhou et al.~\cite{zhou2020ml}, we offered participants a set of initial persona cards, including a parent, a business traveler, a 3D-artist, and an Instagram influencer. For each persona, we listed their goals and photo-taking habits. Motivated by prior work in data-driven design~\cite{helms2018design} and parallel work in data visualization design (i.e., ``data changed everything~\cite{walny2019data}''), we included a set of 15 most recent photos taken by each of the personas. Our data-personas align with the ``minimum-viable-data~\cite{van2018prototyping}'' concept that AI engineers typically use in prototyping machine learning models. From a UX standpoint, the data-driven persona is similar to what designers could generate through user research with mixed-method data (e.g.,~\cite{subramonyam2019affinity}).  We carefully curated the photos to include a variety of images that represented a diverse set of photo capturing behavior and photo content. This was done to ensure a diverse set of AIX solutions.  

\textit{Step 1---Opportunity Spotting ($\sim$25 minutes):} We asked participants to brainstorm ways in which AI can support the decluttering tasks by aligning AI needs with human needs~\cite{PAIR}. We provided them with information about types of AI (predictive, perceptual, generative)~\cite{aimeetsdesign}. We also gave them guidelines about integrating AI experiences into end-user task workflows (e.g., when the AI should automatically perform a task, and when it should take an assistive role when explicitly invoked by end users)~\cite{Appleguidelines}. Participants had access to note pads, sticky notes, and colored markers throughout the session to brainstorm. They were also free to annotate on any of the printed study materials. At the end of this step, participants converged on the AI capabilities they would design in the next steps. 

\textit{Step 2---Model Specification ($\sim$20 minutes):} In this step, participants continued brainstorming ways in which they would implement the AI capabilities that emerged in step 1. We provided them with an ML model design template to brainstorm about training data needs and factors they would consider for implementing the behavior~\cite{aimeetsdesign}. This forced participants to externalize their thoughts and collaborate. We also provided them with printed spreadsheets with persona images on one column and empty columns to fill out with feature values. This encouraged a WoZ like approach to simulate model predictions. They were free to use it as a worksheet to iterate on the design. 

\textit{Step 3---Vertical Prototyping ($\sim$30 minutes):} At this point, we instructed the UX designers to prototype the user interface design using output from steps 1 and 2. We provided them with printed templates for wireframing mobile interfaces. We asked the designers not to use placeholders for text or images in their prototype. We provided them with printed images for each persona (both thumbnail size and screen-size images). Participants could cut and glue the images onto their prototypes (i.e., make medium-fidelity prototypes~\cite{engelberg2002framework}). We intended to see how participants worked with factual data when designing the interface. For text and labels, when they were unclear what the content needs to be, they were asked to annotate with a question mark for later discussion with the engineer. In parallel, we asked the AI practitioner to fill out model API cards (one for each AI capability), providing details about API name, model inputs, model outputs, behavior description, and details about the training data (adapted from ~\cite{mitchell2019model}). This comprised a low-cost realization of the AI ``material'' for feedback and iteration. At the end of this step, they each explained the model API design and the UI design to each other and the study coordinator. We provided participants with translucent sheets (vellum paper) to place on top of the prototypes to annotate and discuss. The goal was to map user interactions to API calls and align model inputs and outputs to the prototype. During this stage, engineers revised the model details through negotiation, and designers updated the interface when required.

\textit{Step 4---Identify AI Errors ($\sim$15 minutes):} We created design cards explaining different types of AI errors and potential sources of errors~\cite{PAIR, amershi2019guidelines}.  Using this information and the prototypes (UI and Model cards), we asked participants to brainstorm AI and UI specific errors for their design. We provided then with a template to document errors along with different categories (system limitation, context error, background error~\cite{PAIR}), but they were free to use the notepad. For this step, participants had to generate a set of potential errors for their AIX design.

\textit{Step 5---Design for Explainability ($\sim$20 minutes):} In this step, we instructed participants to consider explainability as a solution to the errors generated in the previous step. According to guidelines, context errors are a type of AIX errors in which the system is working as intended, but the user might perceive an error due to lack of understanding, or mismatch with their own mental model~\cite{PAIR}. We asked participants to incorporate explainability into their design (both interface prototype and model API) to resolve AI context errors. We provided them with six design cards listing techniques and examples for designing explainable interfaces~\cite{PAIR}. We also provided participants with vellum sheets which they could use to annotate over the prototype to design explainable solutions collaboratively.  

\textit{Step 6---Design for Learnability and Feedback ($\sim$15 minutes):} We asked participants to consider learnability and end-user feedback to improve the model performance. Participants had to design ways to elicit feedback from the users. The key here was to design feedback in a way that can be used for model improvement. We provided participants with information about the types of feedback and guidelines for designing explicit feedback~\cite{PAIR}. 

\textit{Step 7---Setting Expectations for End-Users ($\sim$10 minutes):} In this final step, participants had to design ways to communicate AI capabilities to end-users. We asked them to consider how they might design for end-user trust and how they might design to support end-user control over the data. We provided participants with guidelines about trust and expectation setting~\cite{amershi2019guidelines, PAIR}. Participants could create new wireframes or annotate over existing ones.

Collectively these steps follow the material design process by considering fabrication, application, and appreciation and would allow teams to offer adaptations for AI's uncertainties during the co-creation process. At the end of the study, we debriefed participants about our motivation to investigate AI co-creation process based on HAI guidelines. Participants had the opportunity to ask us questions and provide feedback on the study protocol.

\subsection{Data Analysis}
To prepare the data for analysis, the first author manually transcribed all video recordings. This allowed them to annotate and capture necessary metadata, such as who created the artifacts and how designers and engineers engaged with the study materials. During transcription, they included screen captures of the video to indicate pointing and show-and-tell actions. They also added scanned copies of corresponding artifacts at appropriate points in the transcript. This was done in-line in a word document (one for each session). We then conducted qualitative coding utilizing a combination of deductive and inductive codes~\cite{fereday2006demonstrating}. From literature and our protocol steps, we generated an initial set of codes (e.g., AI fabrication, application design, structure, properties, surface, etc.). For instance, we coded AI implementation details as material structure, and discussions about aligning the UI prototypes and model cards as material surface, i.e., the model API. After coding the transcripts using these codes, we carried out the second round of inductive \textit{in-vivo} coding to analyze the data within each category. The generated codes included references to data and types of communication between designers and engineers (knowledge sharing, negotiation, artifact purpose, validation, guidelines, mentions of UI and AI design elements, etc.). After coding, the authors collectively reviewed and discussed the coded transcripts to identify higher-level themes to answer our research questions on co-creating AIX.

\section{Findings}
We asked participants to co-create an experience for decluttering photo albums using AI (in the abstract). By considering human-centered needs, design guidelines, and user-data context, designers approached the AI material in terms of its \textit{experiential} traits. Engineers, who are technically trained, approached defining AI material in terms of \textit{structural} traits, such as learning algorithms, model features, and model architecture. To bridge these differing viewpoints, teams engaged in rich discussions to ascribe material characteristics to the AI, co-create the application experience and evaluate its fit for end-users (i.e., the user experience). In this process, designers concretized their expectations for the AI material through `designerly' representations, such as scenarios, mental-models, and wireframes. These \textit{shareable} instantiations served as the designers' \textit{proxies} for their desired AI material characteristics. For engineers, these proxies offered human-centered requirements that allowed them to derive the AI material's technical characteristics. We summarize our study findings in terms of (1) \textit{designerly proxies} for articulating AI material needs based on human needs, (2) \textit{data probes} to shape AI material design, and (3) role of \textit{representational artifacts} as realizations of AIX.

\subsection{Designerly Proxies for Articulating AI Material Needs}

\begin{figure*}[t!]
\centering
\includegraphics[width=\textwidth]{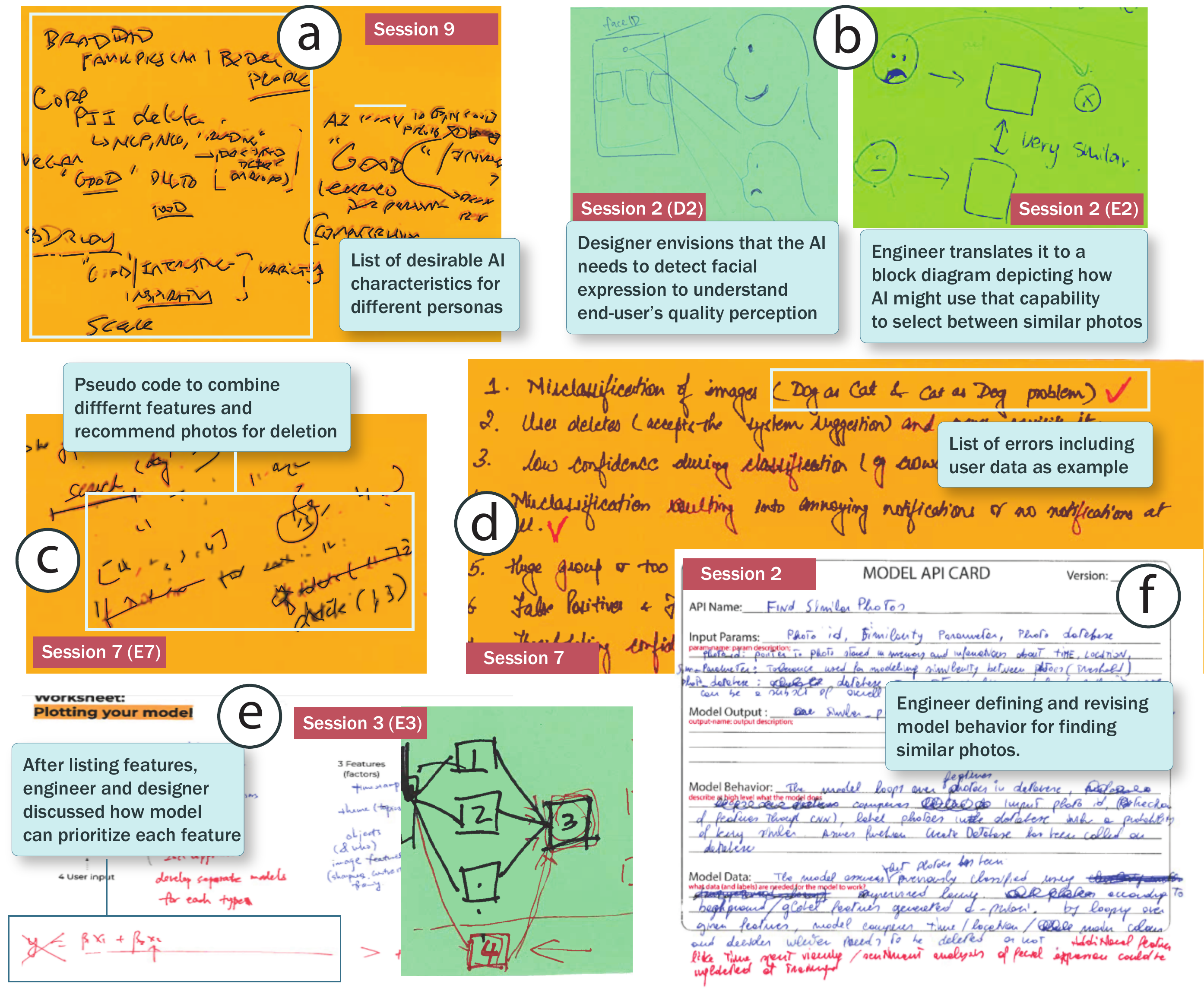}
\caption{Designerly proxies and AI material prototypes created by study participants.}
\Description{Scanned images of participant created artifacts during the study including a scenarios, features, model card, and free-form notes of equations and pseudo code.}
\label{fig:proxies}
\end{figure*}

\subsubsection{Material Properties: User Scenarios as a Proxy for Designing AI Behavior}
Based on the protocol, teams started the design activity by exploring the intersections of user needs and AI strengths. By looking at the personas and their data (photos), the designers constructed different \textit{scenarios} (user vignettes) to identify reasons behind photo clutter. These scenarios captured varied perspectives, including photo-taking (creation context), photo usage, and photos as memory artifacts (archive). For instance, in session 5, the designer (D5) constructed a scenario where a `Dad' persona takes a burst of photos to capture his fidgety kids, intending to keep only the best one. Using such scenarios as anchors, the teams then explored how the AI might support decluttering. The designer then asked the engineer (E5) whether the AI could detect similar photos and identify the best one to keep. This was a \textit{conversational} process consisting of ``thinking out loud'' about different scenarios, supplemented with annotations over the photos (Figure~\ref{fig:teaser} b).  The designer questioned the engineer about AI capabilities: 
\begin{quote}
    D5: `\textit{`Look at what this Dad is doing, he takes lots of photos of his kids and forgets to delete it, so this is one of the main challenges. Can AI identify duplicate photos and find the best one to keep?''}
\end{quote}
\begin{quote}
    E5: \textit{``Yes, we can cluster the images based on similarity\ldots. ''  }  
\end{quote}

By constructing such user scenarios (i.e., material application), designers could ascribe potential capabilities for AI in the abstract and co-create the AI's desirable behavioral characteristics (fabrication). Across all sessions, the scenarios led to instantiations of AI with different behaviors, including parsing text information, image quality assessment, and object recognition (Figure~\ref{fig:proxies}a). Further, in the course of defining AI behavior, designers would revise their initial scenarios to incorporate AI capabilities from engineers, thus creating ``AI-infused'' scenarios. As an example, D4 added to their vignette that the AI could intervene immediately after the person takes the photo: \textit{``After they take photos you wait until they turn off the phone, and then you have a dialog with the user: `Hey these photos, the eyes are shut,'\ldots''}(D4).

\subsubsection{Material Structure: End-User Mental Models as a proxy for Designing AI Implementation}
Design would be incomplete without human-centered considerations about the structure of AI material; that is, \textit{How should the AI do what it is supposed to do?} In step 2 of the protocol, we observed that designers approached this issue by simulating in-depth data walkthroughs with previously defined scenarios~\cite{polson1992cognitive}. During these walkthroughs, the team attempted to construct novel `mental models' about how end-users might make judgments about decluttering their photos (i.e., which to keep and which ones to delete?), and what the AI can be expected to do for them. We call this an \textit{expectation model} of the end-user. For instance, in session 2, the designer considered the moment immediately after taking a photo and the end user's thought process for deciding whether to take a second (Figure~\ref{fig:proxies}b). This led to defining \textit{logic} for whether a photo should be deleted:  \textit{``\ldots the Dad takes a photo of their kids and then views it for three seconds, which means it might be a good photo or a really, really bad photo and they want to improve it\ldots there is some intention behind it\ldots Can you understand from their facial expression [while looking at the photo] whether this is a good photo?''} (D2). 

During these walkthroughs, engineers listened and translated user expectations and decision factors into features, rules, and even pseudo-code for training the model (by writing it down in the model design template - Figure~\ref{fig:proxies}f). They would question designers about the importance of each feature to end-users and how the model can assign different weights to different features. For example, in session 3, the designer first talked about different attributes of a `bad photo.' The engineer then visualized a linear model (See Figure~\ref{fig:proxies}e) to discuss ways to combine those different attributes: 
\begin{quote}
    D3: \textit{``Imagine these four similar pictures, but in this one, I cut his face just a little bit. Can AI identify that? Or if the eyes are closed, and then these are not useful\ldots, and what about lighting or blurriness?''}
\end{quote}
\begin{quote}
    E3: \textit{``From machine learning perspective $y = f(x)$ \ldots and each x can be a feature you put into a small model, and we can aggregate the outputs of each small model into a bigger model\ldots you can run the same picture through each model and weigh the decision from each model whether to delete or not\ldots''}
\end{quote}

From these examples, it is apparent that considering both expectation models and associated technical details helped participants co-create the model's structure. 

\subsubsection{Material Surface: User Interface as a Proxy for Designing AI's API}
The third aspect of the AI material involves how people interact with it through its surface (i.e., the API). For AI, the API drives the user interface for end-user to engage with AI behavior. In our sessions, designers used the interface prototypes as a proxy when co-creating the model APIs. As with scenarios and walkthroughs, prototyping with concrete data points (as opposed to abstract placeholders such as `lorem ipsum\footnote{\url{https://en.wikipedia.org/wiki/Lorem_ipsum}}') allowed designers to articulate specific API-level needs. For example, in session 8, the designer created the interface for viewing a set of recommended photos to delete. By examining this experience, they requested that the API also show key features about \textit{why} the photo was recommended for deletion:  \textit{``Let us think about the workflow, it is time to delete, how do you think they are going to process the photos to delete? Are they going to skim it by looking at the thumbnail, or enlarge it to focus on details? Can you provide a smart thumbnail with the features identified by the AI?''} (D8). In their final design, they proposed `smart-thumbnails' as the AI output. Here, the UI prototypes provided talk-backs for iterating on the material APIs. In most instances, engineers responded by revising the model API card or creating a new API version.
\subsection{Data Probes to Shape AI Material Design}

While designerly proxies offer a medium for articulating AI needs, design also requires generative thinking about alternative solutions. We observed that both the user-data associated with provided personas and participants' `imagined' data (based on their prior knowledge) facilitated generative design processes. Specifically, participants used individual user-data points as \textit{probes} to construct varied designerly proxies, explore the capabilities and limitations of AI, and evaluate created AI material against different HAI guidelines from the protocol. In human-centered design, design probes promote generative thinking and allow designers to explore the design space~\cite{mattelmaki2006design}. We found that end-user data (e.g., the photos associated with data personas) played the role of design probes in crafting the AI experience.

\subsubsection{Data probes for Divergent Thinking}
In constructing user scenarios, participants constantly referred to personas and their photos to brainstorm different AI behavior types. In session 9, referencing the business traveler persona and their photo receipts, the team imagined a natural language understanding behavior to declutter old receipts. As the designer described, \textit{``For this person who is trying to reimburse something, can we delete automatically by reading the text?'' (D9)} E9 responds, \textit{``Yes, we can identify text and what is in it, we can use natural language understanding\ldots''} Besides, the data probes allowed participants to think about different ways to implement AI behavior through \textit{mental models and implementation rules}: 

\begin{quote}
    D7: \textit{``Something that came to me when I was looking at the personas was the emotional connections to these pictures like these pictures have values (pointing at pictures of kids). The application has to acknowledge the value for the users and save them instead. How can you classify pictures that have short term values and those that have long term value?''}
\end{quote}
\begin{quote}
    E7: \textit{``On the back-end, what I hear is that there are different clusters of pictures, and I understand that different pictures have different value\ldots but there is always a possibility of misclassification''}
\end{quote}
\begin{quote}
    D7:\textit{``Could we have overarching rules, like faces might fall into personal attachment bin?''}
\end{quote}

Data probes also allowed participants to explore various surface-level features like explainability and end-user feedback to the model. For example, in session 6, the designer did not want to display confidence scores to end-users. To this, the engineer used example data points to illustrate why a lack of explanation may result in distrust for end-users: \textit{``Would the user be displeased if they took four photos and all four had bad lighting, but the system showed the photo with the highest score as this is the best photo because of lighting, but in actuality, they are all bad\ldots Would that cause a loss in trust?'' (E6)} Then the designer agreed \textit{``That is a good point, maybe we have 4-5 categories [features], and you show a score underneath for each \ldots, so you know that all four photos are bad.'' (D6)}. Similarly, in all sessions, participants used data probes to determine the type of feedback that might be useful for model improvements. In session 3, the engineer explained that binary feedback of whether the recommendation was good or bad might not be sufficient for the model to improve. According to the engineer, \textit{``How to use feedback is a difficult task for ML\ldots the thing is users deciding to keep an image may not just be because the prediction is wrong, but because they just like it\ldots for example in this picture the eyes are closed, but she looks cute\ldots One way to design the interface is to have more than two buttons; in addition to keep or delete, if there could be a third button, like yes, eyes closed, but I still want it\ldots''(E3).} 

\subsubsection{Data Probes for Exploring Material Boundaries and Limitations}
A key aspect of designing AI is understanding the ``edge cases,'' where the AI might fail, or the designed behavior may not work. This understanding is essential for selecting the best alternative at design time (i.e., maximize potential for material appreciation) and designing for uncertainty during use, including failure, error handling, explainability, learnability, and setting expectations. In many instances, designers used data probes to decide whether to incorporate specific AI behaviors into the design. For instance, in session 4, the engineer suggested using smile detection to determine `goodness' of the photo. To this, the designer pushed back by commenting:\textit{``\ldots if they had a stroke and are not smiling [in the picture], deleting that would be bad\ldots''} (D4). In this case, they incorporated their understanding of the broader domain to think about examples beyond our data personas.  

We also observed several instances in which engineers used example data points to highlight the limitations of AI. For instance, E4 commented, \textit{``The model might predict that she is not smiling because of a missing tooth, or braces\ldots''} E5 cautioned about AI limitations on classifying images of kids: \textit{``For adults, facial recognition works well, but I am not sure whether it will work for kids. These all may be photos of the same kids, but I mean kids grow\ldots the clustering may not work.''}. These examples helped participants determine whether the user experience with AI should be automated or assistive: \textit{``For duplicate photos, I think it should be assistive and complementary\ldots if the user deletes one photo, we can say here are other photos like this, would you like to delete those as well?''} (E10). 

\subsubsection{Data Probes for Design Evaluation}
In addition to exploring material qualities, data probes also supported design convergence and confirmatory evaluation. Similar to identifying edge cases, participants made use of data probes to check whether a behavior `scaled' across different personas. For example, in session 10, participants identified the desired AI behavior of detecting duplicates and selecting the best photo to keep. The designer then considered an Instagrammer persona for testing whether the same behavior might be useful to them. According to D10: \textit{``Same goes here\ldots after they upload to Instagram they probably do not need it. Especially for Instagram people, they edit their photos a lot, and each time they edit, it will create new copies if it\ldots''}. Similarly, after making decisions about \textit{not} including certain behaviors in their design, teams discussed why that decision was right by considering other personas. For instance, in session 9, they rejected their initial idea about predicting images' temporal utility. In acknowledging their decision, the engineer E9 commented: \textit{``I am somewhat nervous these [frequency] metrics\ldots for instance, a big failure in which case is deleting a baby's birth picture, that would be bad\ldots''}
\subsection{Representational Artifacts as Realizations of the AI Material}

\begin{figure*}[t!]
\centering
\includegraphics[width=\textwidth]{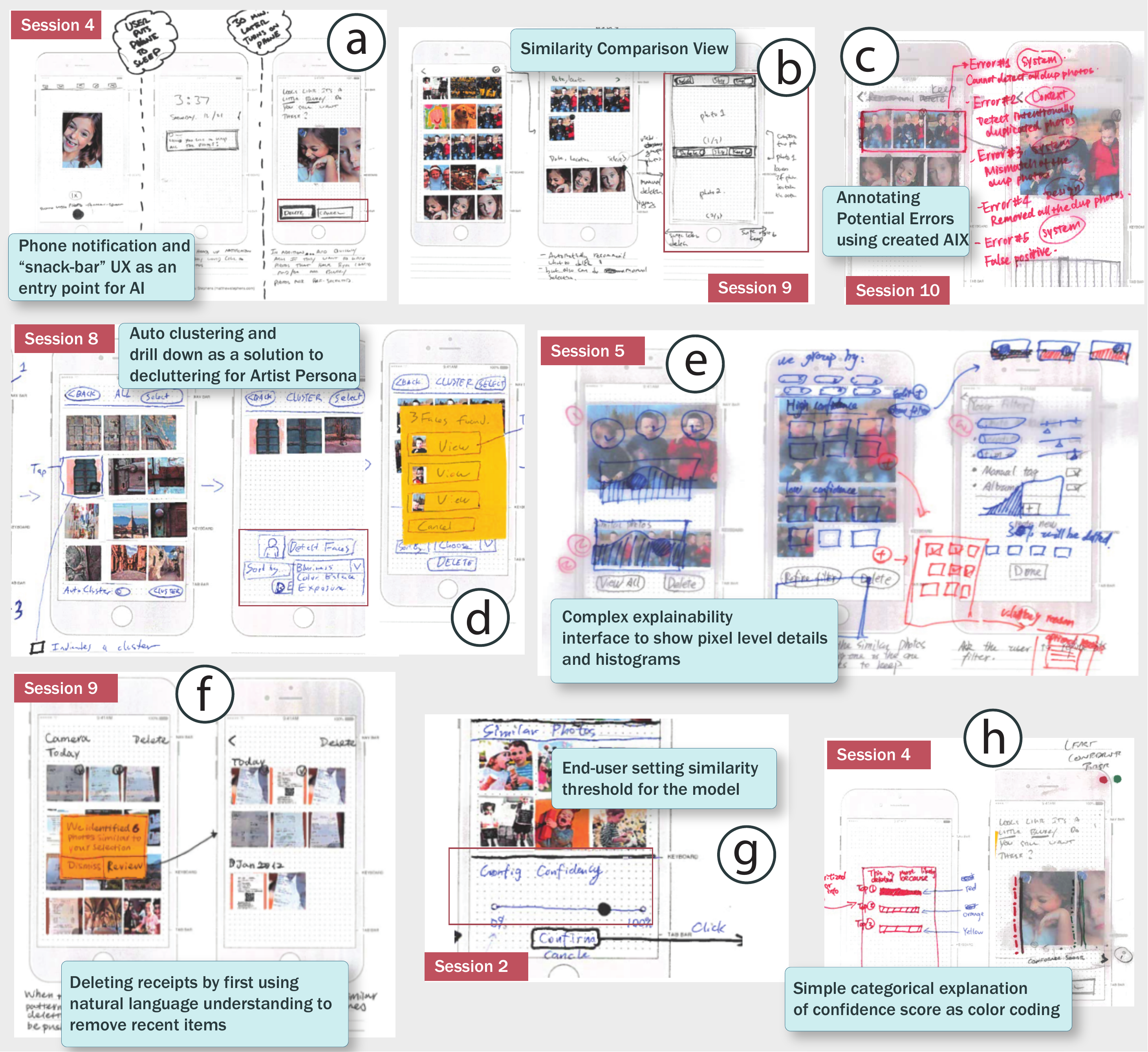}
\caption{AIX Representational Artifacts showing interface design solutions including explainability, and model feedback.}
\Description{Scanned images of UI prototypes and annotations over vellum paper}
\label{fig:wireframes}
\end{figure*}

Prototyping with user-data allowed designers to experience the AIX `first-hand' as they were creating it---material appreciation. It also allowed them to concretely communicate their use of the created AI material back to the engineers by ``representing it to make the solution transparent~\cite{simon1969sciences}''. Given the simplicity of the design brief, we were surprised to see the wealth of design variations generated across the sessions. As shown in Figure~\ref{fig:wireframes}, the entry points for the declutter UX included explicitly invoking the AI via a button click, use of an AI-triggered notification, end-user (delete) actions, providing seed images as a starting point, and conversational dialog. Participants also designed various AI-powered presentation views, including intelligent clustered lists, comparison views highlighting feature differences, interactive thumbnail views, etc. Across all prototypes, the selection of images was driven by imagining AI-infused scenarios and using it to `play-out' the experience as it was created. Here, we discuss three key benefits to AI material prototyping, including: (1) aligning AI and UI through translation and feedback, (2) addressing misconceptions and gaps in understandings about AI, and (3) helping designers perceive the complex nature of the AI material. 

\subsubsection{Aligning AI and UI through Translation and Feedback:}
In constructing AIX prototypes, designers had to translate their understanding of the AI model's structure and APIs into their own knowledge of user interface design. Drawing from their design expertise, they began by representing both the AI behavior and output through familiar design patterns. In many sessions, this provided a starting point for envisioning AIX. In session 2, to recommend to users which photos to delete, the designer started with a familiar photo album interface: \textit{``How to separate good one from all the junk ones? In Apple, they show you all the photos, and you select which ones you want to keep or not\ldots'' (D2)}. Similarly, in session 4, the designer started with their familiarity about `snack-bar' UI (i.e., a notification with quick action buttons) to prototype an `assistive' AI experience: \textit{``\ldots they have what is called a snack bar\ldots where it is a confirmation message, but within the confirmation message you can have a like an undo button, that is really common now\ldots'' (D4)}. 

During the prototyping process, designers concretized their understanding of the model by considering model input and feedback controls within the context of user-data. Their choice of text (such as labels and dialog) corresponded with the AI model implementation rules and features; e.g., in session 4 prototype: ``it looks like it's a little blurry, do you want to keep this?''. We also observed that designers intuitively designed certain presentation features that led to team discussions and subsequent changes to the model API. In session 4, the designer D4 decided to bin confidence scores into higher-level categories and color code them in photos (Figure~\ref{fig:wireframes}h). Their rationale was to make it more accessible to end-users who may not understand the meaning of differences in confidence scores. This design decision prompted a discussion with the engineer about how the model might categorize confidence scores for a more intuitive presentation:
\begin{quote}
    D4: \textit{``you can give them like some sort of color highlight, so you can give them a threshold\ldots''}
\end{quote}
\begin{quote}
    E4: \textit{``That's a good point, you can order them by confidence, and you can give them a threshold\ldots we can test it out by having a training dataset and test dataset, and we can say from our average, a typical use case the most evenly divided percentages are for example anything below 50 is low, 50 to 75\% is medium, anything above is high confidence.''}
\end{quote}
The material prototype allowed the team to anticipate user needs and redesign the AI material to address them. 

\subsubsection{Addressing Misconceptions and Gaps in Understanding}
Representational prototypes allowed both designers and engineers to identify gaps in each others' understanding. Through discussions and annotation overlays (using vellum paper), they negotiated differences in their understanding of the AI material. For instance, in session 8, the designer had prototyped detailed text explanations about why a set of images were clustered together. Looking at the prototype, the engineer commented: \textit{``I am not sure technically machines are capable of that\ldots people are good at generating semantic explanations, for example people can tell others in natural language why these photos are similar or why they are clustered together\ldots But like I am not sure even state of the art ML models are capable of that'' (E8)}. By contrast, in session 9, the AI engineer suggested using personalized sorting algorithms to present the photos to be deleted. The designer then annotated onto the prototype to clarify that sorting should be objective, and subjective sorting would likely make the user not trust the recommendations. 

\begin{quote}
    E9: \textit{``We can start from a fixed equation and adjust parameters based on user interactions\ldots, for example, this image has more exposure than others, then we can personalize the sorting algorithm by changing parameters.''}
\end{quote}
\begin{quote}
    D9:\textit{``This is supposed to be an objective way of sorting the value, so if I knew that this sorts blurriness [order photos by blur level] to my choice, I would not trust it anymore.''} 
\end{quote}

\subsubsection{Perceiving AI Complexities}

The act of constructing an AI material prototype made designers more aware of the complexities of AI-powered interfaces. Across many sessions, by looking at material flaws (e.g.,  material uncertainties that do not communicate rationale to end-users), designers discovered additional needs, such as setting default model parameters, feedback features, explainability, setting expectations, etc. Working through several iterations on their design, the designer in session 6 commented: \textit{`` Now we are `frankensteining' this sucker, but you can have a settings page that pops open over here and allows them to say, `keep top N photos'\ldots ''}. 

While the co-creation process clarified the underlying material structure for designers, some found it hard to separate from their increased knowledge about the AI in order to design the UX, or to communicate their AI understanding to end-users through their design. We observed instances in which the user-to-AI feedback mechanism became very complicated, reflecting the designer's understanding of the AI model but failing to use that understanding in designing the UX. For example, in session 5, the designer and engineer engaged in rich discussions about categorization failures based on pixel-level features and personalization. In applying this new understanding, the designer created a very complicated user interface prototype without considering trade-offs between improving AI model accuracy and the user effort required for feedback (see Figure~\ref{fig:wireframes}e). These examples demonstrate changes in conceptualizations of AI material arising through the co-design of AIX.

\section{Discussion}
Design materials are central to design processes. In conventional UX design, the graphical user interface (GUI) is the prevalent ``design material'' that every UI designer understands~\cite{carroll1988interface, fernaeus2012material}. If AI were like any other material defined by nature or convention, designers would learn how to work with its given properties to generate design solutions for human users. However, AI resists this approach. Had we provided participants a `created' AI material (e.g., a closed-box ML model that assigned a quality score to each image~\cite{talebi2018nima}), they likely would not have produced the range of expressive, human-centered designs that we observed. Instead, as our findings show, AI material must be defined by investigating the human user's envisioned experiences. In our study, we aimed to answer: (1) How do designers and engineers conceptualize design guidelines from AI and UX perspectives? (2) How do they co-create the design and technical characteristics of AI materials; and (3) What representations are invented during this process? Based on our findings, we respond to these questions by proposing a process model for co-creating AIX and by reflecting on the role of end-user data as a design probe for generative design thinking. Through this discussion, we offer design considerations for data probes within AIX design tools. 

\subsection{Towards a Process Model for Co-Creating AIX}

\begin{figure*}[t!]
\centering
\includegraphics[width=\textwidth]{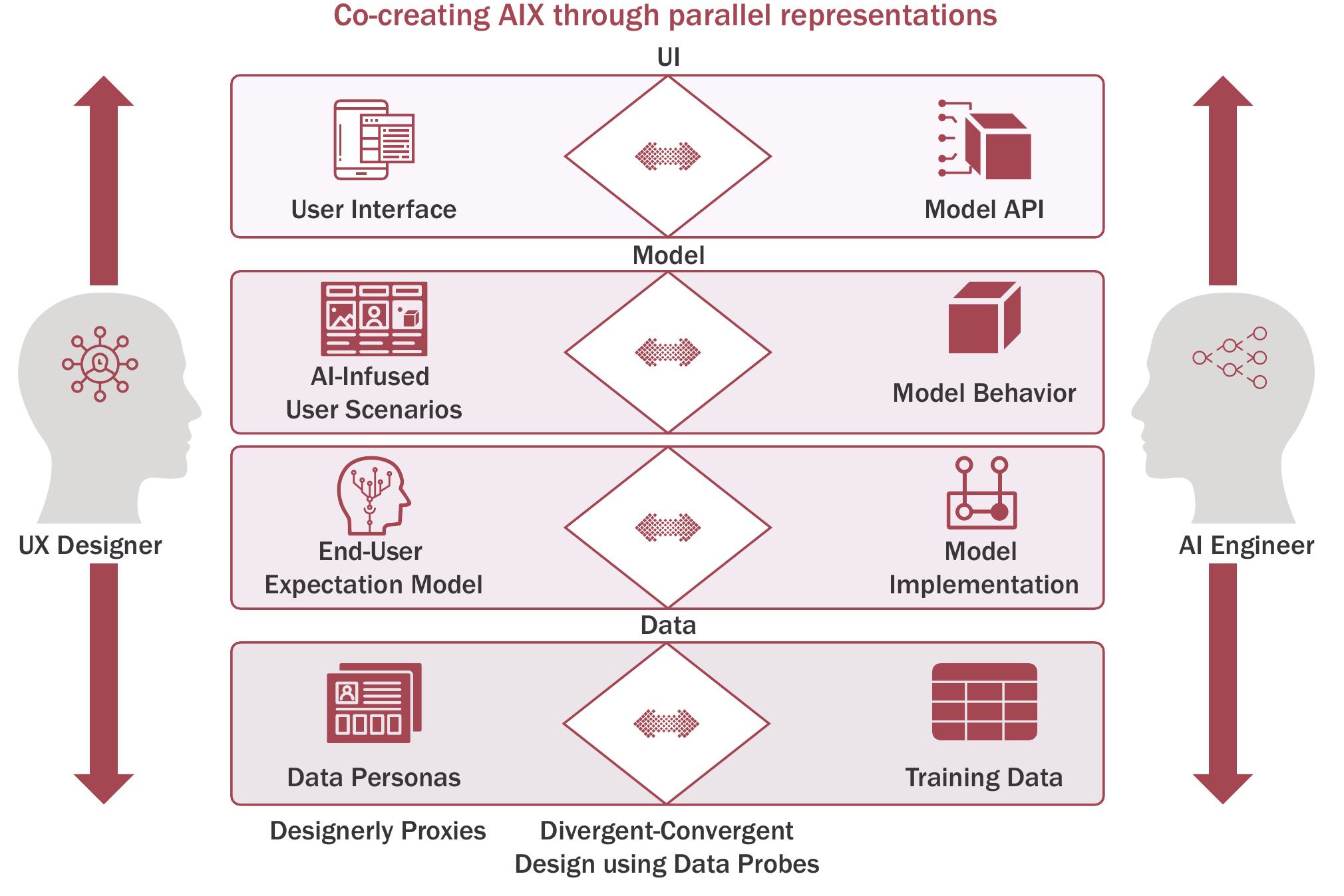}
\caption{A process model for co-creating AIX.}
\Description{An illustration of the parallel representations used by designers and engineers to co-create the AI material and application experience.}
\label{fig:processmodel}
\end{figure*}

In the study, our teams displayed a design progression occurring across a \textit{spectrum} of materiality. Teams started the design activity by exploring the intersections of user needs and AI strengths. First, with wholly \textit{imaginary and abstract} material, the team worked together to envision the AI. The designers negotiated this initial form-giving by constructing user scenarios, allowing them to approach AI through its potential capabilities without detailed engineering knowledge. This is similar to the role of visualization as proxies in designing `immaterial' materials~\cite{arnall2014exploring}. Designers used personas, data points, vignettes, and user scenarios---their designerly proxies---to create initial instantiations of the desired AI behavior. These proxies offered design representations as abstractions of \textit{planned behaviors} that allowed engineers to define the technical characteristics of the AI material. In the course of co-designing AI behavior, designers also revised their initial  scenarios to incorporate new AI capabilities, thus creating ``AI-infused'' scenarios.

Equipped with an invented form, teams moved forward to \textit{enact and specify} the AI material. Following human-centered walkthroughs of scenarios, the designers constructed novel \textit{expectation models} capturing how the AI's end-users might make judgments. The engineers translated these identified user expectations and decision factors into features and rules for training the AI. This co-creation process led to discussions about the attributes, priorities, and values important to users and the technical capabilities the AI needed to support them. While the AI material has taken on behavior and structural characteristics, its envisioned design was only fully \textit{realized} through the team's use of material prototypes. Designers used interface prototypes as a proxy when co-designing the model's inputs and outputs (i.e., the material's surface). These designerly proxies allowed the team to align the AI and UI through translation and feedback. Identifying specific AI and human behaviors allowed the evaluation of material flaws, misconceptions, and scalability issues. This is similar to \textit{Replay Enactments}~\cite{holstein2020replay} that use authentic data to make complex system behavior tangible to designers. Only at this late stage could the full scope of the AI material and UX design be made visible in its interactive complexity.  

Clearly, the \textit{dynamic} nature of AI material is unlike other design materials; consequently, the design process for AIX differs from standard design approaches. In conventional human-centered design (HCD), like the double diamond framework~\cite{council2005double}, the design process is linear or top-down. Designers mainly work at the user interface layer to specify the end-user experience. They then hand-off the created specifications to engineers to build~\cite{seffah2005human}. However, when designing AI experiences, design extends \textit{beyond} the interface and into the design of AI components, including the model's behavior, learning characteristics, assumptions, and nature of training data. UX professionals lack the means to engage in designing these AI components. Instead, current AI development workflows take an ``AI-first'' approach in which the AI material is created before envisioning its use. Such an approach is problematic because any changes to align the AI's properties to human needs will require costly rework (e.g., addressing disparities in gender classification~\cite{buolamwini2018gender}). Further, there are instances in which the AI behavior itself does not align with human needs, values, and concerns (e.g., using facial recognition to expose political orientation~\cite{kosinski2021facial}). 

In order to address these issues, we need a process in which AI material creation and its application experience design can happen in parallel through iteration and feedback. As shown in Figure~\ref{fig:processmodel}, we propose a process model that combines top-down (UX-first) and bottom-up (AI-first) workflows to distribute agency between designers and engineers. As represented by the \textit{bidirectional} arrows in our model, the AI and UX components are designed in parallel, a critical insight from our study. Our approach shifts engineers' mindsets towards more proactive engagement through accessible user-data proxies and data probes during the co-creation process. Designers engage in co-creating AI behavior without technical roadblocks, operationalize HAI guidelines, and reduce time to feedback (a concern with AI design~\cite{yang2019sketching}). Our model's parallel process affords immediate feedback for both material creation and design, obviating the significant rework costs (from collecting and training with new training dataset, retraining the models, etc.) when even small changes arise later. However, we also emphasize the ``towards'' in the paper's title; Our study lays groundwork for a collaborative process to align AI's form and its function in the early stages of design. Future research should build on this parallel process model to investigate specific data and representation needs across different application domains and AI capabilities.

\subsection{Role of Data Probes in AIX Design}
Data probes served as a ``content common ground'' for designers and engineers to collaborate across the different stages in the process model. A characterization of this collaboration is that designers are immediate consumers of AI materials. Their objective is to ensure that the material specification meets their end-users' UX needs.  Using data probes, designers advocated for end-users during the material creation process and simultaneously tested the AI material under construction. Similar to~\cite{register2020learning}, end-user data offered necessary grounding for designers to advocate for centering people in the design of AI, including its behavior, implementation, and APIs.

As shown in Figure~\ref{fig:processmodel}, each of the parallel stages in our model involved both divergent and convergent processes. Designers and engineers ideated on UX needs and AI capabilities together, and they mutually constrained convergence towards a design solution. User data as probes played a critical role in this divergent-convergent process of creating the AI experience. We can extend the material metaphor and borrow from the language of physical material design to characterize the role of data probes: (1) data \textit{molds}, (2) data \textit{vulcanizers}, and (3) data \textit{coupons}. In the early stage of the study, data probes functioned as ``molds'' for AI's initial form-giving. By constructing AI-infused scenarios with data probes, designers and engineers explored different forms the AI could potentially take in supporting the declutter experiences. After identifying the initial form, designers used data probes to define the AI material's internal properties. By constructing expectation models with data probes, designers and engineers ``solidified'' the AI's implementation requirements. This step in AI material creation is analogous to `vulcanization' chemicals in the rubber manufacturing process to solidify its internal structure. Finally, by constructing AIX interface representations with real data, designers produced coupons (test samples) of the material to assess the AI experience. In traditional material design processes, `coupons' are samples of the material used to test its properties at a small scale (e.g.,~\cite{huang2014art}). The designers' mixed-fidelity prototypes served as coupons to test the AI material and address gaps in the desired AI experience.

\subsection{Design Considerations for AIX Design Tools}
Prototyping is an essential step in software development~\cite{seffah2005human}. Through iterative prototyping, teams incorporate increasing details to define different software aspects ~\cite{beaudouin2009prototyping}. The mixed-fidelity approach in our study is an initial step to iterative prototyping. As details increase, teams need to increase the fidelity of their prototypes as well. In this regard, UX design tools should escape the ``closed-box'' view to make AI more accessible and transparent to designers ~\cite{yang2018investigating}. Beyond helping designers understand AI (i.e., educational goals), designers should be able to work with AI material during AIX design. The insights from our study suggest that data probes offer useful design considerations for this goal. With this in mind, we offer a set of design considerations for incorporating data probes into design tools.

\textit{Support for creating data probes:} In the current study, we constructed the data for each persona to include a variety of solution alternatives. Participants also imagined their own additional data points during the design process. Design tools should allow designers to incorporate data from user research into AIX design processes directly. This could include data collected from participants (similar to Wizard-of-Oz prototyping~\cite{browne2019wizard}), through mixed methods persona creation (e.g., Data-Assisted Affinity Diagramming~\cite{subramonyam2019affinity}), or from dedicated data collection and annotation pipelines~\cite{girardin2017user}. In addition, tools should support accessible ways to generate user data with desired properties. In the data visualization community, tools exist to create datasets with desirable statistical properties (e.g.,~\cite{mannino2019real, Drawmydata}), allowing designers to select charts to fit their data needs. Our study teams imagined varied data---blurry images, variations in size, and time-progression photos---using their understanding of the task set within use contexts. Design tools should support such expressive `queries' to find or generate \textit{just-in-time} data probes for designers. 

\textit{Support for interactive AI \& UX design workflows:} To work with the AI material under construction, designers can use data probes to receive ``talk-back'' from the AI design workflow. Currently, end-user machine learning tools such as RunwayML~\cite{runwayml} allow novices to interact with machine learning models and for visual exploration of machine learning behavior (e.g., the What-If Tool~\cite{whatIf}). However, these tools do not support the use of data probes for divergent thinking about AI behavior. To be effective, probes should be integrated into generative prototyping workflows to provide input and feedback to designers. Moreover, interactions using data probes allow designers to propose desired outputs based on human needs. This can be in the form of ground truth data, annotated labels, output format, etc. For instance, designers can curate a set of diverse data points and ground truth outputs and compare the working model's output against ground-truth values. 

\textit{Support for constructing designerly proxies:}  Currently, when prototyping UIs, designers typically work with static placeholder content. As with data visualization, in AI, ``data changes everything~\cite{walny2019data}.'' Prototyping tools should allow designers to construct AIX design candidate representations by incorporating data probes and AI material talk-backs (e.g.,~\cite{klemmer2000suede}). This allows consideration of alternative choices~\cite{maclean1991questions}, along with design for AI uncertainty through explainability, learnability, and edge-case  analysis\cite{amershi2019guidelines}. 

\textit{Support for communication during co-creation:} A final consideration for co-creating AIX is to share intermediate proxies of AIX, including scenarios, mental models, and interface prototypes with AI engineers. For engineers, they need to offer descriptions of AI properties, assumptions,  learning rules, and API details back to designers. This is fundamentally different from standard workflows in which designers primarily share final design specifications with engineers. Both UX design tools and AI creation tools should incorporate features to import, translate, and share intermediate design proxies. Similar to the transparent vellum paper during our study, digital tools should support annotation overlays, generate new examples with different data probes, and communicate failures and constraints (i.e., explainability for designers). 

\subsection{Limitations and Future Work}
Our study's design problem was to define an AI-powered experience for decluttering photo albums. While this simple problem allowed us to observe co-creation processes in an accessible domain, other data types may be more challenging for co-creation. More complex design problems may be difficult to represent with pen-and-paper approaches. We plan to conduct co-design sessions with other data types and problem domains to iterate on the process model. We aim to assess (and address) the fit and shortcomings of our process model for different AIX problems through these sessions. Second, we provided participants the data probes for use in our study. We plan to investigate inclusive and participatory approaches to creating diverse data probes for AIX design. Third, our protocol demonstrates a low-cost, rapid prototyping approach to co-creating AIX. As described, this is the first step to iterative prototyping with increasing levels of fidelity. We are currently exploring high-fidelity prototyping tools for AIX to understand how teams might continue to evolve their designs. As an example, we developed ProtoAI~\cite{subramonyam2021protoai} to allow designers to invoke models and services with concrete data during prototyping. Fourth, we recruited participants with prior experience in designing AI applications. In many industries, both designers and engineers are new to AI \cite{cai2019software}. We are now using the study protocol in classrooms to teach design and engineering students about AIX design. Through these efforts, we will investigate the types of training and scaffolding designers need to effectively participate in AIX's rapid prototyping. Finally, future work should investigate how other stakeholders, including domain experts, representative end-users, and data analysts, might participate in the AIX co-creation process. 

\section{Conclusion}
Treating technology as a design material encourages designers to explore its properties for UX design. However, when working with AI as design material, neither a form-follows-function nor a function-follows-form approach is practical. Instead, the AI material and its application UX need to be co-created through collaboration between designers and AI engineers. In this work, we investigated such an approach by conducting an in-lab design study with ten pairs of designers and engineers. Our protocol combines a vertical prototyping approach with talk-backs from AI and UX to facilitate co-creation.  We identified the crucial role of end-user data as a tool for co-creating AI design material. By using data probes, designers were able to construct designerly proxies and specify material needs for AI. Data probes facilitated divergent thinking, material testing, and design validation. Based on these findings, we propose a process model for collaborative AIX design and offer considerations for incorporating data probes in AIX design tools.

\begin{acks}
We thank our reviewers and study participants for their time and feedback. We also thank Ken Holstein, Steven Drucker, and Steve Oney for their inputs on the paper. 
\end{acks}

 \bibliographystyle{ACM-Reference-Format}
 \bibliography{99_refs}

\end{document}